\documentstyle[12pt]{article}

\centerline{\bf { A Remark 0n the Geometry of Two-Dimensional}} 
\centerline{\bf { Anisotropic Non-Linear Sigma-Models}}
\vspace{0.7cm} 
\centerline{D. H. T. Franco${^{(*)}}$\footnote{dfranco@cbpf.br},  M. S.
G\'oes-Negr\~ao${^{(*)\dag}}$\footnote{negrao@gft.ucp.br}} 
\centerline{ J. A. Helay\"el-Neto${^{(*)\dag}}$\footnote{helayel@gft.ucp.br} and A. R. Pereira${^{(+)}}$}
\vspace{.5cm} 

\centerline{$^{*}$Universidade Cat\'olica de Petr\'opolis(UCP-GFT)} 
\centerline{$^{\dag}$Centro Brasileiro de Pesquisas F\'{\i}sicas(CBPF-MCT) } 
\centerline{$^{(+)}$ Universidade Federal de Vi\c{c}osa} 
\vspace{.5cm}

\begin{document}

\hspace\parindent
{
\begin{abstract}
\vspace{0.3cm}

{\footnotesize {\bf One discusses here the connection between
$\sigma$-model gauge a\-nom\-a\-lies and the existence of a connection
with torsion that does {\underline{not}} flattten the Ricci tensor of the target
manifold, by considering a number of non-symmetric coset spaces. The influence of an
eventual anisotropy on a certain direction of the target manifold is also contemplated.}}
\end{abstract}}

Pacs numbers: 02.40.-k; 11.10.-z; 11.10.Lm.
\vspace{.5cm}
\newpage

\section{Introduction}
\hspace\parindent

{ The geometrical setting, the classical dynamics and
quantum-mechanical properties of non-linear $\sigma$-models have
been extensively discussed by field-theorists over the past decade.
Phenomenological motivations connected to low-energy effective
physics \cite{1} and considerations of a more formal nature,
regarding ultraviolet behaviour, renormalisability \cite{2} and
all-order finiteness of some classes of $\sigma$-models \cite{3}
have been the leitmotif for pursuing a thorough investigation of
these non-linear theories.

Introducing supersymmetry, a number of issues that concern the
formulation of  $\sigma$-models are brought about \cite{4}. One of
the outstanding points that comes out has to do with the partner fermions
that, by means of supersymmetry, naturally couple to the  bosonic
coordinates of the target space \cite{4a}.

The dynamics of fermions coupled to a $\sigma$-model requires a
vector bundle, $B$, with a connection defined over the target
space, $M$, on which the bosonic fields, $\varphi^{i}$, map the
space-time manifold. The coupling to eventual by present chiral fermions,
as dictated by supersymmetry for instance, to the connection naturally defined
on $B$ is a potential source of anomalies, as clearly discussed in
the works of ref. \cite{5}.

Our paper sets out to analyse how it might happen that possible
geometrical constraints may be imposed  on the target space
geometry, so as to prevent the appearance of anomalies connected to
the gauge fields associated to the connection introduced in B, or
to the isometry gauge fields coming in whenever one performs
the gauging of isometries (or isotropies) of $\sigma$-models defined
on homogeneous spaces of the type $G/H$.

If $dim$ $G/H = D$, the anomaly alluded to in ref. \cite{5} is the
one associated to the $SO(D)$ -- pull-back connection of $B$. In
our case, we shall take into account the situation where one
introduces extra gauge fields that come into play with the purpose
of gauging the subgroup $H$ (suitably embedded in $SO(D)$) or even
the whole isometry group, $G$.

This problem has been previously tackled in a series of works
\cite{6}, and the connection between the geometrical structure of
$G/H$, and the mechanism for cancellation of the isometry-group
gauge anomaly has been worked out in a paper by Alvarez- Gaum\'e
and Ginsparg \cite{7}, where the authors succeed in fixing
conditions on the $H$-subgroup content of the fermion fields in such a way
to eliminate the isometry-group anomaly. Our purpose here is to
understand if there is a relationship between the $H$-subgroup
attributes of the fermions and the torsion on $G/H$, in such a way
to characterize the anomaly suppression mechanism more directly in
terms of the geometrical structure of the target space. The main
motivation behind our attempt is the coupling between the torsion
of the $\sigma$-model manifold and the fermion bilinears. We choose
to consider here non-symmetric coset spaces  with non-vanishing
torsion, having in mind that interesting geometrical properties may
arise at the expenses of working with a metric connection with
torsion. We also try to exploit the geometric nature of anisotropic
non-linear $\sigma$-models. Topological aspects of the latter
have been considered by Watanabe and Otsu \cite{wat} who have shown that anisotropy may lad to non-trivial metastable states
that generate local minima of the energy (instantons).

\section{Anisotropic non-linear $\sigma$-models and non-symmetric
spaces}
\hspace\parindent

Following the work of ref. \cite{8}, we shall specify the geometry
of $G/H$ in terms of group-theoretical properties of $G$ and $H$.
First of all, we assume the splitting dictated by the decomposition
\begin{eqnarray}
adj \ G &=& adj \ H \bigoplus V,
:\nonumber\\ \label{1}[Q_{i}, Q_{j}] &=& {f_{ij}}^{k} Q_{k}, \nonumber\\
\label{2.a} [Q_{i}, Q_{a}] &=& {f_{ia}}^{b} Q_{b}, \nonumber\\ \label{2.b}
[Q_{a}, Q_{b}] &=& {f_{ab}}^{i} Q_{i} + {f_{ab}}^{c} Q_{c}
,\label{2.c}
\end{eqnarray}
where the $Q_{i}$'s ($i$ = $1,..., $ $dim$ $H$) denote the
generators of $H$ and $V$ refers to the $Q_{a}$'s ($a$ = $1,...,D$)
the generators of $G$ outside $H$.

Denoting by $\varphi^{\alpha}$ the coordinates of a point in the
coset $G/H$ ($\alpha$ is the world index and runs from $1$ to $D$),
the {\it vielbein} and {\it connection} are obtained by means of
the following $G$-Lie-algebra-valued one-form:
\begin{eqnarray}
e(\varphi) {\equiv} L_{\varphi}^{-1} dL_{\varphi} =
[e_{\alpha}^{i}(\varphi)Q_{i} + e_{\alpha}^{a}(\varphi)Q_{a}]
d{\varphi}^{\alpha},
\label{3}
\end{eqnarray}
where $L_{\varphi}$ is a coset representative and $a$ = $1,...,D$
is to be identified with the tangent space index (local frame
index).

With the help of the Cartan-Maurer equation for the 2-form $d
e(\varphi)$, 
\begin{eqnarray}
de(\varphi) =  -e(\varphi) \wedge e(\varphi),
\label{4}
\end{eqnarray}
with $ e \equiv e^{i}Q_{i} + e^{a} Q_{a}$, the relevant geometrical objects
(torsion 2-form, $T^{a}$, and curvature 2-form, ${R^{a}}_{b}$)
that obey the equations,
\begin{eqnarray}
d e^{a} + {w^{a}}_{b} \wedge e^{b} = T^{a}
\label{5.a}
\end{eqnarray}
and
\begin{eqnarray}
{R^{a}}_{b} = d{w^{a}}_{b} + {w^{a}}_{c} \wedge {w^{c}}_{b},
\label{5.b}
\end{eqnarray}
turn out to read in components as follows:
\begin{eqnarray}
{w^{a}}_{b} = -{f^{a}}_{bi} e^{i} - \frac{1}{i} {f^{a}}_{bc} e^{c}
-
\frac{1}{i} {T^{a}}_{bc} e^{c}.
\label{6}
\end{eqnarray}
If one chooses the torsion 2-form to be given as below,
\begin{eqnarray}
T^{a} = \frac{1}{2} k {f^{a}}_{bc} e^{b}
\wedge e^{c},
\label{7}
\end{eqnarray}
$k$ being an arbitrary coefficient, then the connection read as follows:
\begin{eqnarray}
{w^{a}}_{b} = - {f^{a}}_{bi} e^{i} - (1 + k){f^{a}}_{bc} e^{c};
\label{8}
\end{eqnarray}
this, in turn, yields
\begin{eqnarray}
{R^{a}}_{bcd} &=& {f^{a}}_{bi} {f^{i}}_{cd} +
\frac{1}{c} (1 + k){f^{a}}_{be} {f^{e}}_{cd} +
\nonumber\\
&+& \frac{1}{4} (1 + k^{2}) ({f^{a}}_{ce}{f^{e}}_{db}
- {f^{a}}_{de}{f^{e}}_{cb}).
\label{9}
\end{eqnarray}
Torsion shall be absent if $k = 0$, or in the case the embedding of
$H$ into $G$ is a symmetric one.

Anisotropic non-linear $\sigma$-models are, as ordinary $\sigma$-models, theories of maps
between manifolds. More precisely, the scalar fields $\phi^{i}$ of
the theory map a given space-time, X, to a given target space, M.
The action of model is obtainable from a pure kinetic term
$\frac{1}{2}[\sum_{i}{\partial_{\mu}}{\varphi^{i}}{\partial^{\mu}}{\varphi^{i}}
+
{\lambda}{\partial_{\mu}}{\varphi^{k}}{\partial^{\mu}}{\varphi^{k}}]$
(i=1,2,...,k,...n), by solving the constraint
${\sum_{i}}{\varphi^{i}\varphi^{i}}=1$, to obtain
\begin{eqnarray}
S=\frac{1}{2} \int
d^{2}xg_{ij}(\varphi){\partial_{\mu}}{\varphi^{i}}{\partial^{\mu}}{\varphi^{j}},
\label{acao}
\end{eqnarray}
with
\begin{eqnarray}
g_{ij}(\varphi) = {\delta}_{ij} +
(1+\lambda)\frac{\varphi_{i}\varphi_{j}}{1-{\varphi}^{2}},
\label{met}
\end{eqnarray}
where $i,j \neq k$ and ${\lambda}>-1$. The meaning of ${\lambda}$ shall be clarifield below.

By rewriting the original fields as ${\tilde\varphi}= \biggl (
{\varphi}_{1}, {\varphi}_{2},..., (1 + \lambda)^{1/2}{\varphi}_{k},
..., {\varphi}_{n} \biggl )$, we get the surface $S^{n-1}$ on an
n-dimensional spheroid,
\begin{eqnarray}
\sum_{i \neq k} {\tilde \varphi}^{i}{\tilde \varphi}^{i} + \frac{{\tilde \varphi}^{k}{\tilde \varphi}^{k}}{b^{2}} = 1,
\label{sphe}
\end{eqnarray}
where $b^{2}=(1+\lambda)$. From eqs.(\ref{met}) and (\ref{sphe}),
we see that the anisotropy parameter, $\lambda$, deforms the usual
metric on a sphere.

In two dimensions, the action(\ref{acao}) is not the most general
one. Namely, assume that the target space carries, besides the
given metric $g$, a given two-form, $\omega$. Then, the complete
action is given by
\begin{eqnarray}
S = \frac{1}{2} \int d^{2} x g_{ij}(\varphi) {\partial}_{\mu}
{\varphi}^{i}{\partial}^{\mu}{\varphi}^{j} + \frac{1}{2} \int
d^{2}x {\epsilon}^{\mu \nu} {\omega}_{ij}(\varphi) {\partial}_{\mu}
{\varphi}^{i} {\partial}_{\nu} {\varphi}^{j},
\label{acao2}
\end{eqnarray}
where we refer to ${\omega_{ij}}(\varphi)$ as the torsion
potential, since it is well-known that this term introduces torsion on
the manifold \cite{muk}.

We shall consider the following ansatz for the torsion:
\begin{eqnarray}
T^{a} = \frac{1}{2}(1+\lambda)kf^{a}_{bc} e^{b}\wedge e^{c},
\end{eqnarray}
where $k$ is related to the torsion degree of freedom and $(1+\lambda)$ gives
the dependence of the torsion on the anisotropy parameter. It
then follows that
\begin{eqnarray}
{\omega}^{a}_{b}=-f^{a}_{bi}e^{i} - \frac{1}{2}
(1+k+k\lambda)f^{a}_{bc}e^{c}.
\end{eqnarray}
This allows us to obtain the curvature and the Ricci tensor,
respectively,
\begin{eqnarray}
R^{a}_{bcd}=f^{a}_{bi}f^{i}_{cd} +
\frac{1}{2}(1+k+k\lambda)f^{a}_{be}f^{e}_{cd} +
\frac{1}{4}(1+k+k\lambda)^{2}(f^{a}_{ec}f^{e}_{bd}
- f^{a}_{ed}f^{a}_{bc}),
\end{eqnarray}
\begin{eqnarray}
R_{ab}= f^{c}_{ai}f^{i}_{cb} +
\frac{1}{4}[1-(1+\lambda)^{2}k^{2}]f^{c}_{ae}f^{e}_{cb}.
\end{eqnarray}

Now, our main idea is to try to understand how the fermion coupling
to the torsion of $G/H$ may probe the anomaly matching condition
proposed in the work of ref.\cite{7}. Since we have not yet found
a general line of arguments that lead to find the geometrical
counterpart of the condition quoted above, it is our idea to set
the geometry of some non-symmetric spaces, namely,
$Sp(4)/U(1)\times U(1)$, $SU(2)\times SU(2)/U(1)$, $G_{2}/[SU(2)]^{2}$, $G_{2}/SU(3)$,
$SU(3)/U(1)\times U(1)$ and $Sp(4)/(SU(2)\times U(1))_{nonmax}$, to
illustrate that the {\underline{anomaly cancellation}} condition found in ref.\cite{7} has to do with the {\underline{non}}- {\underline{possibility of finding a
Ricci-flattening connection with torsion}}.

For the latter three homogeneous spaces listed above, it is always
possible to have a Ricci-flattening connection, whenever the
coefficient $k$ in eq.(\ref{7}) and the anisotropy coefficient
fulfill the condition:
\begin{eqnarray}
(1+\lambda)k = \pm \sqrt{5}.
\label{10}
\end{eqnarray}
This is so because
\begin{eqnarray}
R_{ab} = {f^{c}}_{ai}{f^{i}}_{cb} + \frac{1}{4} (1 -(1+\lambda)^{2}
k^{2}){f^{c}}_{ad}{f^{d}}_{cb} \label{11}
\end{eqnarray}
can be taken vanishing by virtue of the results below \cite{guida}
\begin{eqnarray}
{f^{c}}_{ai}{f^{i}}_{cb} = {f^{c}}_{ad}{f^{d}}_{cb} = \left \{
{\begin{array}{lll} \delta_{ab}, for \hspace{0.3cm}
SU(3)/U(1)\times U(1); \\ \delta_{ab}, for \hspace{0.3cm}
Sp(4)/(SU(2)\times U(1))_{nonmax}; \\ \frac{4}{3}\delta_{ab}, for
\hspace{0.3cm} G_{2}/SU(3).
\end{array}} \right.
\label{12}
\end{eqnarray}.

However, for $Sp(4)/U(1)\times U(1)$, $SU(2)\times SU(2)/U(1)$ and $G/[SU(2)]^{2}$,
Ricci-flatness {\underline{cannot}} be achieved at the expenses of torsion. For
$SU(2)\times SU(2)/U(1)$, this is so because $rank((SU(2)\times
SU(2))/U(1))\neq 1$; nevertheless, even though $rank$$Sp(4)=rank
(U(1)\times U(1))$ and $rank$$G_{2}=rank
(SU(2)\times SU(2))$, it is {\underline{not}} possible to find a solution for $k$
that enables to set $R_{ab}=0$. This result can be understood with
the help of the explicit calculations of the combination of
structure constants of the isometry groups related to these spaces. For example, in the case of $Sp(4)$, explicit calculations yield \cite{guida}:
\begin{eqnarray}
{f^{c}}_{ai}{f^{i}}_{cb} = \left
\{{\begin{array}{lll} 1, for \hspace{0.3cm} a=b=2,4,5,8,9; \\ 2,
for \hspace{0.3cm} a=b=3,6,7; \\ 0, for \hspace{0.3cm} a\neq b
\end{array}} \right.
\label{13}
\end{eqnarray}
whereas
\begin{eqnarray}
{f^{d}}_{ac}{f^{c}}_{db} =
\left \{ {\begin{array}{lll} 2, for \hspace{0.3cm} a=b=3,4,6,7; \\
4,  for \hspace{0.3cm} a=b=2,5,8,9; \\ 0, for \hspace{0.3cm} a\neq
b.
\end{array}} \right.
\label{14}
\end{eqnarray}
Analogously, calculations carried out for $G_{2}$ \cite{guida} also show that {\underline{no}} $k$ exists that enables us to set $R_{ab}=0$. 

What then remains is to analyse the connection between the
non-sym\-me\-try of the space and the non-existence of a Ricci-flattening
connection with torsion on the other hand, the fulfilment of the anomaly
cancellation condition on the other hand, as expressed in terms of the H-content of
the fermions \cite{7}\cite{mauro}.

Our claim, by now only supported by explicit examples and not by a
general approach, is that, if $rank$ $G = rank$ $H$, and if {\it torsion
is non-trivial and does not allow the vanishing of the Ricci
tensor}, then the isometry and/or isotropy group anomaly does {\it not} show up. In
the cases torsion flattens the manifold (in the sense it yields a vanishing Ricci tensor), one notices that
the anomalies are {\it not} cancelled, as it is the case for
$SU(3)/U(1)\times U(1)$, $Sp(4)/(SU(2)\times U(1))_{nonmax}$ and $G_{2}/[SU(2)]^{2}$.

For (complex) projective spaces like $CP^{n}$ and Grassmannians, which
appear in the framework of $N=1-D=4$ and $N=2-D=2$ supersymmetric
${\sigma}$-models, isometry anomalies persist as there is no
torsion, once these are all symmetric spaces.


\begin{thebibliography}{99}

\bibitem{1} S. Coleman, J. Wess, and B. Zumino, {\em Phys. Rev.} {\bf 177}
(1969) 2239; C. G. Callan,Jr., S. Coleman, J. Wess and B. Zumino,
{\it ibid.} {\bf 177} (1969) 2247; E. Cremmer, S. Ferrrara, L.
Girardello and A. Van  Proyen, {\em Phys. Let.} {\bf 116 B} (1982)
231 and {\em Nucl. Phys} {\bf B 212} (1983) 413.

\bibitem{2} D. Friedan, {\em Phys. Rev. Let.} {\bf 45} (1980) 1057; L.
Alvarez-Gaum\'e and D. Z. Freedman, {\em Phys. Let.} {\bf 94 B}
(1980) 17; Alvarez-Gaum\'e, D. Z. Freedman and S. Mukhi, {\em Ann.
Phys.} {\bf 134} (1981) 85.

\bibitem{3} Alvarez-Gaum\'e and P. Ginsparg, {\it ``A Class of
Two-dimensional Finite Field Theories''} in proceedings of
Symposium on Anomalies, Geometry and Topology, edited by W. Bardeen
and A. White, World Scietific, 1985.

\bibitem{4} B. Zumino, {\em Phys. Let.} {\bf 57 B} (1979) 203.

\bibitem{4a} Kiyoshi Higashijima and Manueto Nitta, {\bf hep-th/0008240}.

\bibitem{5} G. Moore and P. Nelson, {\em Phys. Rev. Let.} {\bf 53} (1984)
1519.

\bibitem{6} J. Bagger and E. Witten, {\em Phys. Let.} {\bf 118 B} (1982)
103; J. Bagger, {\em Nucl. Phys.} {\bf B 211} (1983) 302; C. M.
Hull, A. Karlhede, U. Lindstr\"om and M. Rocek, {\em Nucl. Phys.}
{\bf B 266} (1986) 1.

\bibitem{wat} T. Watanabe and H. Otsu, {\em Prog. Theor. Phys.} {\bf 65} (1981) 164.

\bibitem{muk} S. Mukhi, ``Non-linear Sigma Model, scale invariance and string theories: A pedagogical review'', Summer Workshop in High Energy Physics and Cosmology, ICTP, Trieste, 1986.

\bibitem{8} A. Salam and J. Strathdee, {\em Ann. of
Phys.} {\bf 141} (1982) 316.


\bibitem{7} Alvarez-Gaum\'e and P. Ginsparg, {\em Nucl. Phys.} {\bf B 262}
(1985) 439.

\bibitem{mauro} M. S. G\'oes-Negr\~ao, {\it M.Sc.Thesis} ``Anomalies on Nonlinear Sigma Models", Institute of Physics , UFRJ, Rio de Janeiro.

\bibitem{guida} P. Brockill, J. A. Helay\"el-Neto and M. R. Negr\~ao, ``Group Theory for Field Theorists", Lectures Notes, CBPF, Rio de Janeiro.


\end{thebibliography}
\end{document}